\documentclass[iop]{emulateapj}
\usepackage{epsfig}
\usepackage{graphics}
\usepackage{graphicx}

\newcommand{\msun}{$M_\odot$}
\newcommand{\lsun}{$L_\odot$}

\shorttitle{L1634 IRS~7 Variability}
\shortauthors{Hodapp}

\begin{document}

\title{Periodic Accretion Instabilities in the Protostar L1634~IRS~7\\}

\author{Klaus~W. Hodapp\altaffilmark{1},
Rolf~Chini\altaffilmark{2}$^,$\altaffilmark{3} }

\altaffiltext{1}{
Institute for Astronomy, University of Hawaii,
640 N. Aohoku Place, Hilo, HI 96720, USA
\\email: {\em hodapp@ifa.hawaii.edu} }

\altaffiltext{2}{
Astronomisches Institut, Ruhr-Universit{\"a}t Bochum,
Universit{\"a}tsstra{\ss}e 150, D-44801 Bochum, Germany
\\email: {\em rolf.chini@astro.ruhr-uni-bochum.de}
}

\altaffiltext{2}{
Instituto de Astronomia,
Universidad Catolica del Norte,
Avenida Angamos 0610, Antofagasta, Chile
}

\begin{abstract} 
The small molecular cloud Lynds 1634 contains at least three outflow sources.
We found one of these, IRS~7, to be variable with a period of $37.14 \pm 0.04$ days and
an amplitude of approximately 2 mag in the $K_s$ band. The light curve consists
of a quiescent phase with little or no variation, and a rapid outburst phase.
During the outburst phase, the rapid brightness variation generates light echoes
that propagate into the surrounding molecular cloud, allowing a measurement of the
distance to IRS~7 of 404 pc $\pm$ 35 pc. We observed only a marginally significant
change in the $H - K$ color during the outburst phase. The $K$-band spectrum of IRS~7 
shows CO bandhead emission but 
its equivalent width does not change significantly with the phase of the light curve.
The H$_2$ 1--0 S(1) line emission does not follow the variability of the continuum
flux. We also used the imaging data for a proper motion study of the
outflows originating from the IRS~7 and the FIR source
IRAS 05173-0555, and confirm that these
are indeed distinct outflows.
\end{abstract}


\section{INTRODUCTION}

The Lynds 1634 dark cloud is a small molecular core in the general proximity of the
Orion Molecular Cloud (OMC) that shows a bright rim in H${\alpha}$, indicating the presence of ionizing
UV radiation. Core mass measurements are in the range of one or two dozen \msun, e.g.,
\citet{DeVries2002} measured 12 \msun~in N$_2$H$^+$ and 28 \msun~in HCO$^+$, while
\citet{Beltran2002} measured 3 - 10 \msun.
The L1634 core contains at least three objects in the protostellar outflow phase. 

The first outflow to be discovered was the system of
Herbig-Haro objects 240. 
The brightest knot of this system of shock fronts
is optically detectable as RNO~40
and its proper motion away from the main embedded far infrared (FIR) source -- later called IRAS~05173-0555, but 
originally discovered by \citet{Cohen1985} -- in
a generally western direction was
determined from photographic plates by
\citet{Jones1984}.
The western outflow lobe is blueshifted, while the eastern lobe is redshifted \citep{Lee2000}.

The driving source of this spectacular outflow, the far-infrared source IRAS~05173-0555, has a 
luminosity of 17 \lsun~measured by \citet{Reipurth1993}. There has been some debate as to whether
the IRAS source should be classified as SED Class 0 or 1, summarized, e.g., in \citet{Beltran2002}. 
Its spectral energy distribution,
sub-mm luminosity, and powerful outflow, 
as well as the infall signature observed at millimeter wavelengths would support
classification as Class 0, while the fact that the central object is detectable at near infrared
wavelengths argues for Class I status. In any case, it is clear that IRAS
05173-0555 
is a very young object
with an estimated age of $\approx$ 10$^5$ yrs,
still
accreting substantially and driving a bright outflow with multiple shock fronts. 

The near-infrared emission of the shock fronts associated with the L1634
IRAS 05173-0555 outflow was first
seen in the $K^{\prime}$ survey by \citet{Hodapp1994} and then studied in detail 
in H$_2$ 1--0 S(1) line emission by \citet{Hodapp1995} and \citet{Davis1997}.
\citet{OConnell2004} presented an imaging and spectroscopic study of the 
western half of the HH 240/241 outflow associated with IRAS~05173-0555, concluding that
the H$_2$ 1--0 S(1) emission is excited in internal C-shocks within the outflow.
In their 1995 paper, Hodapp \& Ladd identified a second, 
less luminous outflow source, labeled IRS~7, and
two S(1) shock fronts plausibly associated with it that they labeled as objects 4 and 9. 
The FIR map by \citet{Cohen1985} had already indicated an extension of the emission to the east of
the main FIR source, and the authors at the time attributed this to emission from an HH object, but
in retrospect, that extension indicated the presence of IRS~7.
When compared to the IRAS~05173-0555 source, IRS~7 is readily detectable in the near-infrared and thus less obscured, 
its sub-mm luminosity is lower,
and its well-collimated outflow is less powerful.
All this indicates that IRS~7 may be slightly more evolved than the FIR IRAS source and may be a typical
Class I object in its late accretion phase.

\begin{figure}
\begin{center}
\figurenum{1}
\includegraphics[scale=0.37,angle=0]{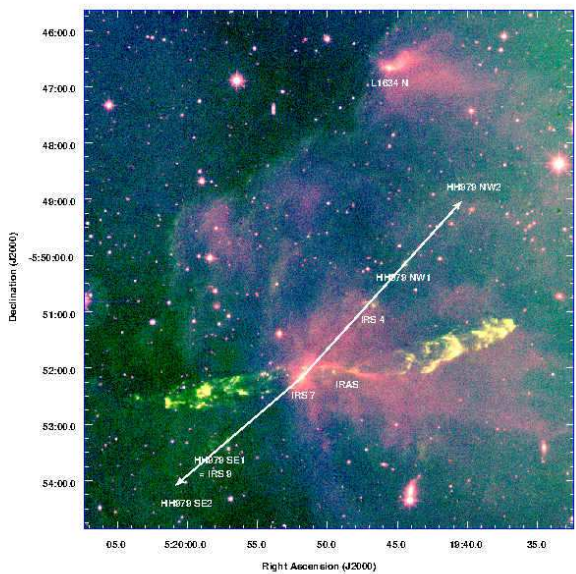}
\caption{
Infrared color composite image of the L1634 cloud: Blue is the $H$ band, green is the S(1) line image, and red is
the $K$ band. The two outflow sources 
IRS~7 and IRAS 05173-0555
are clearly visible by their redder color, while H$_2$ shock fronts appear green. The edge of the molecular cloud  
is faintly outlined by 
S(1) emission (green), probably fluorescently excited.
The axes of the IRS~7 outflow lobes are indicated.
}
\end{center}
\end{figure}

In their comprehensive study of the optical emission in L1634 
\citet{Bally2009} identified optical
[SII] emission knots along the same axis as defined by IRS 4, 7, and 9 of \citet{Hodapp1995} and 
labeled these as components of the 
HH~979 system of Herbig-Haro objects.
In Fig.~1, we present an overview color composite image of L1634, based on
our $H$, S(1) narrow band, and $K$ images, with the objects found by
\citet{Hodapp1995} and \citet{Bally2009} labeled 
and the axes of the IRS~7 outflow indicated.
The next system of HH objects downwind to the NW of IRS 4 was labeled
HH~979~NW1 by \citet{Bally2009}. In the other outflow lobe, the \citet{Bally2009} HH~979~SE1
system of S[II] emission knots is identical to the \citet{Hodapp1995} IRS~9 system. The fact that most
shock fronts of the IRS~7 outflow are optically detectable confirms again that this system 
is associated with less dense molecular material and is probably more evolved than IRAS 05173-0555
and its outflows.

More detailed sub-mm maps were published by \citet{Chini1997} 
and \citet{Beltran2002}, clearly showing some sub-mm emission from the IRS~7 source. 
While \citet{Beltran2008} have mapped the outflow cavity of the main (IRAS 05173-0555) flow,
no NH$_3$(1,1) emission was detected at the position of IRS~7.
\citet{Bally2009} identify a third outflow, labeled L1634N, which is located at the northern edge of the
L1634 cloud. On our images, we do not detect any shock-excited H$_2$ 1--0 S(1) emission at that position.

\citet{Bally2009} have assumed a distance of 400 pc, on the basis 
of the best distance determinations to the Orion Nebula cluster
of 389 pc based on VLA parallax measurements by \citet{Sandstrom2007}
and 392 pc from the rotational properties of young stars by \citet{Jeffries2007}
and the argument that
the bright rim of L1634 is likely illuminated by
the Orion OB1 association stars. 

This paper reports the discovery that the L1634 IRS~7 outflow source
is rapidly and periodically variable.
We measure a distance of L1634 of 404 pc, consistent with the arguments above.
We also present proper motion measurements of the H$_2$ shock fronts over a time span of 19 years,
providing convincing kinematic evidence that IRS~7 is indeed 
the source of the outflow traced by IRS 4, IRS 9, and HH 979.

\section{OBSERVATIONS}

\subsection{IRIS Discovery of the Variability of L1634 IRS~7 and UKIRT follow-up}

As part of an on-going program of monitoring star-forming regions for variability,
the area of L1634 was repeatedly imaged in the $K_s$ band with the IRIS 0.8m telescope
and 1K 2.5~$\mu$m infrared camera
\citep{Hodapp2010} 
of the Universit\"atssternwarte Bochum 
on Cerro Armazones, 
operated jointly by the 
Ruhr Universit\"at Bochum, Germany and the Universidad Catolica del Norte in
Antofagasta, Chile.
For each visit of a specific region, IRIS takes 20 raw data frames on the object with an
integration time of 20~s, with small dithers
to get around detector artifacts, followed by 20 dithered data frames on a separate sky field
chosen to avoid bright stars or extended emission. For these sky exposures, larger dithers
are used than for the object exposures to minimize the impact of any extended emission.

\begin{figure}
\begin{center}
\figurenum{2}
\includegraphics[scale=0.5,angle=-90]{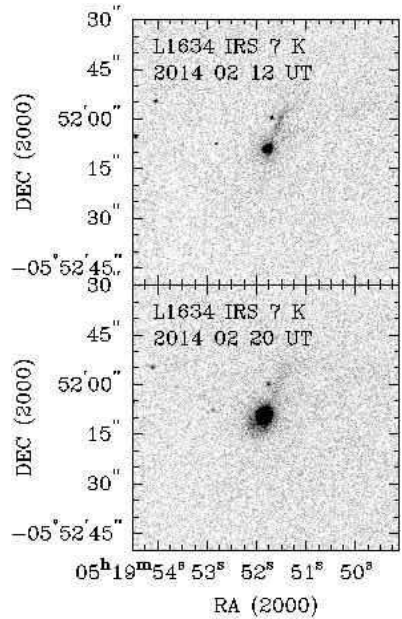}
\caption{
Two UKIRT images of L1634 IRS 7 near minimum and maximum brightness. The images were taken 8 days apart.
}
\end{center}
\end{figure}

After variability with a high amplitude and short time constants
had been found in the near-infrared object IRS~7,
we conducted a dedicated monitoring campaign, using the IRIS system and UKIRT with WFCAM
\citep{Casali2007}
as often as was practical
in the observing seasons 2013-2014 and 2014-2015.

The UKIRT observation with WFCAM were done with an individual integration time of 10~s,
a dither pattern of 9 jitters, and separate sky exposures with the same parameters.
In total, 27 UKIRT images of
good quality were obtained in the $K$ band.
Two of the resulting UKIRT images are shown in Fig.~2, and show IRS~7 in the minimum phase and
in the rising phase of the light curve.
The data from both IRIS and UKIRT are shown in the light curve Fig.~3.
We have also observed L1634~IRS~7
using WFCAM on UKIRT in the $H$ filter, and in the narrow-band S(1) line filter.
The $H$-band images were taken within minutes of the $K_s$ images and were therefore
suitable for a measurement of the object color. 
The S(1) line images formed the second epoch data set for a study of the shock-front proper motions.

\begin{figure}
\begin{center}
\figurenum{3}
\includegraphics[scale=0.35,angle=-90]{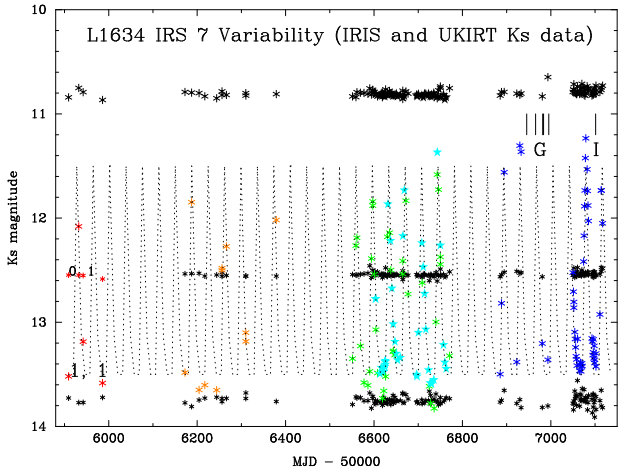}
\caption{
L1634~IRS~7 $K_s$ light curve from IRIS and UKIRT photometry. 
Different observing seasons are shown in different colors, 
and the same color coding is used in the phased light curve in Fig.~5.
The black star symbols show the photometry of non-variable field
stars of different brightness.
The epochs of the Gemini (G) and IRTF (I) spectroscopic observations are indicated.
}
\end{center}
\end{figure}

\subsection{Gemini GNIRS and IRTF SpeX Spectroscopy}
In the observing season 2014 to 2015, we obtained spectra under Gemini project GN-2014B-Q-90
with the Gemini Near-Infrared Spectrograph (GNIRS) \citep{Elias2006} in 5 nights, using a 0$\farcs$45
wide slit centered on IRS~7. The cross-dispersed mode was used, giving, in principle, the full
spectrum from 1.2$-$2.5~$\mu$m. However, in L1634 IRS~7, only the $K$-band emission was bright enough
for spectroscopy. 
The GNIRS spectrograph slit was oriented at P.A. 145$^{\circ}$, aligned with the direction of the outflow
of IRS~7 given by the two emission knots IRS~4 and 9.
In each of the 5 nights, six individual spectra of IRS~7 were obtained, interleaved with six spectra
of the night sky a few arcseconds away from IRS~7 and any extended emission associated with it
or the other shock fronts. Each spectrum consisted of six co-added exposures of 10~s integration time.

One additional spectrum was obtained on March 20, 2015, 
with the SpeX instrument \citep{Rayner2003} at the Infrared Telescope Facility (IRTF). 
For these observations, a 0$\farcs$8 wide slit was used that
covered much more of the extended S(1) line emission around IRS~7 and was oriented at P.A. 160$^{\circ}$
aligned with the outflow features close to the central object IRS~7. The SpeX observations consisted of
11 on-object frames of 120 s integration time, interleaved with the same number of sky exposures. Since these observations were 
taken with a wider slit and longer integration time than the Gemini/GNIRS spectra, the extended S(1) line
emission contributes more to this spectrum and the S/N of the continuum is comparable, even though the IRTF
is the smaller telescope.

\section{Data Reduction}

\subsection{IRIS Data Reduction Pipeline}
The IRIS camera raw data were processed in a reduction pipeline based on
the Image Reduction and Analysis Facility (IRAF) software \citep{Tody1986}.
The raw data were flatfielded using incandescent light dome flats.
Sky frames were obtained at positions near the object specifically selected 
to be free of bright stars and extended emission. 
The optical distortion correction was based on an as-built model of the optical
system. After correcting the known optical distortions, the
astrometric solutions of the individual images were calculated based on SExtractor \citep{Bertin1996} and
SCamp \citep{Bertin2005} based on 2MASS coordinates \citep{Skrutskie2006}. 
The images were then co-added using these astrometric solutions.
Aperture photometry was obtained with the IRAF task PHOT with an aperture of $r = 4$ pixels ($r = 3\arcsec$) 
and calibrated against an ensemble of stars from the 2MASS
catalog. These 2MASS reference stars were selected for proper brightness,  
being isolated from other stars and from nebular emission. 
For part of the monitoring period, the IRIS telescope experienced unusually large pointing errors and
unstable image quality due to a mirror support problem, so that
there are only 9 suitable 2MASS photometric reference stars common to all IRIS frames used here. 
The 2MASS reference stars were in the brightness range 12.5 - 14.5.
For the IRIS data the rms scatter of the measured magnitudes ranged from
0.03 for the brightest to 0.10 mag for the faintest reference stars,
while for the UKIRT data, the rms variations were 0.007 for the brightest
and 0.026 for the faintest reference stars.
These rms uncertainties are consistent with those measured on larger
samples of reference stars in other fields, and do not indicate any
significant variability of any of the reference stars.
In Fig.~3, we show the measured data of three stars in the field,
two reference stars and one other, bright, apparently constant star that was
not used as a reference star because its image overlaps with a background
galaxy labeled VLA1 in \citet{Bally2009}, and its photometry quality flags
therefore did not meet our criteria. The brighter of the reference stars
has a magnitude standard deviation of 0.007 for UKIRT and 0.027 mag for IRIS,
the fainter one has a magnitude rms of 0.013 for UKIRT and 0.040 for IRIS.

\subsection{UKIRT Photometry}
The UKIRT data were processed by the Cambridge data reduction pipeline. We used the data in the
form of fully reduced and co-added images, and extracted the photometry from those images using
the same procedures and reference 2MASS stars as were used for the IRIS data.
For consistency, we also used the same large photometric aperture as for the IRIS data, even though
the image quality of the UKIRT data would have allowed much smaller apertures to be used. For the study
of the relatively bright IRS~7, the resulting increase in background noise was irrelevant.
The UKIRT WFCAM uses a K-band filter based on the Mauna Kea Observatory (MKO) standard described by \citet{Tokunaga2002},
and the color transformation between this WFCAM $K$-band and the 2MASS photometry has been
given by \citet{Hodgkin2009}. Our 2MASS reference stars have an average $J - K$ color of 1.0 mag, while
for IRS~7, the 2MASS catalog gives $J - K$ = 4.5 mag. The small but significant color correction of 
0.035 mag has been applied to the UKIRT photometry of IRS~7, so that all photometry in Figs. 3 and 5 is in the 2MASS $K_s$ system.
However, this subtle correction has no bearing on any of our conclusions.

\subsection{Period Determination and Light Curve}

From all the $K_s$ photometric data over the time span from 2012 to 2015, obtained either with IRIS or
UKIRT, a period of the variability of 
$37.14 \pm 0.04$ days 
was determined (Fig.~4), using the Lomb-Scargle
algorithm developed by \citet{Lomb1976} and \citet{Scargle1982}, 
and implemented in the Period Analysis Software (PERANSO) written by T. Vanmunster. 
In addition to the Lomb-Scargle algorithm, PERANSO offers twelve other period determination
algorithms. We used all of these on our data set, and used the 
scatter of these independent period determinations as a measure of the error associated
with the Lomb-Scargle period value. From this method we estimate that the period of
L1634~IRS7 has an rms error of 0.04 days, i.e., approximately one hour.

\begin{figure}
\begin{center}
\figurenum{4}
\includegraphics[scale=0.35,angle=-90]{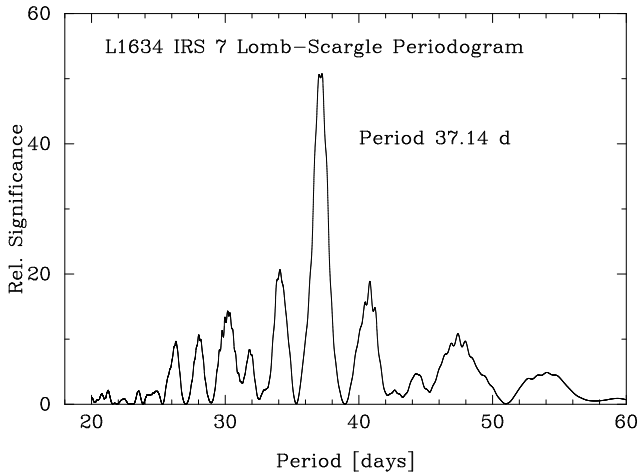}
\caption{
Lomb-Scargle periodogram of the combined IRIS and UKIRT $K_s$ photometric data on L1634 IRS 7.
A period of 37.14 days is clearly indicated. 
}
\end{center}
\end{figure}

The light-curve resulting from folding all our data with the 37.14 day period is displayed
in Fig.~5. with the maximum light at phase 0.5.
L1634 spends about half the time, from phase 0.8 to 0.3, near its minimum brightness,
the other half of the period is an outburst with a brightness increase between one and two magnitudes
in the $K_s$ band. The brightness increase starts at phase 0.30 and the maximum is at
phase 0.50. 
The declining phase of the light curve can roughly be divided into a fast decline, from phase
0.5 to about 0.8, and then a slow, residual decline from phase 0.8 to approximately, and
coincidentially, phase 1.0. A Kolmogorov-Smirnov comparison of the data in the phase range 0.8 to 1.0
to the data in 
the quiescent phase from 0.0 to 0.3 gave a 4$\sigma$ confidence that these are not randomly drawn from
the same population.

\begin{figure}
\begin{center}
\figurenum{5}
\includegraphics[scale=0.35,angle=-90]{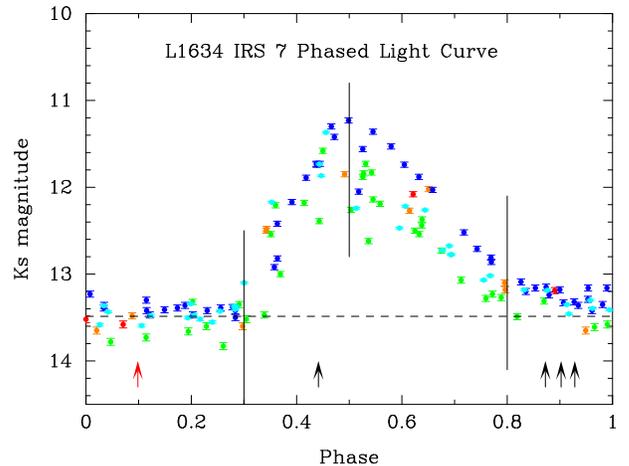}
\caption{
Lightcurve of L1634 IRS~7 phased with a period of 37.14 days determined from all IRIS and UKIRT data points
by the Lomb-Scargle algorithm.
The scatter in magnitude is much larger than the measured photometric errors and therefore indicates variations
in both the maximum and minimum brightness from one period to the next. 
The arrows indicate the phases when the GNIRS (black) and SpeX (red) spectra were taken.
}
\end{center}
\end{figure}

From the best sampled observed maximum, the highest of all photometric
values obtained was at MJD~57079.1145 with a $K_s$ = 11.23 magnitude. In the phased light curve
(Fig. 5), this data point was at phase 0.4988. We estimate an uncertainty of the timing
of this maximum of 0.1 d. From this we derive the following
ephemeris for the maxima of L1634 IRS~7: MJD $(57079.16 \pm 0.1) + N \cdot (37.14 \pm 0.04)$.

The photometric errors were determined from the scatter of the measurements on constant
stars in the field around L1634~IRS~7, and are indicated as error bars in Figs. 3 and 5.
These photometric errors, 0.01 mag rms for UKIRT and 0.03 mag for IRIS at the magnitude
level of the outburst, are much smaller than the variations observed in the light curve.
We conclude that the quiescent-phase brightness of L1634~IRS~7 varies within a range of
$\approx$~0.5~mag, and the peak brightness varies by as much as a full magnitude from
outburst to outburst.

In the last observing season 2014-2015 (blue points in Fig.~5), the well-sampled light curve lies above all
previously recorded data points, both in the quiescent phases and during the outbursts.
By comparison, in the 2013-2014 season (green points in Fig.~5), the IRIS and UKIRT data points were
below most other data points. This indicates variations
in outburst amplitude, but the data from the earlier years do not indicate a secular brightnening trend.

\subsection{UKIRT Data Phased Images}
At every epoch the WFCAM data in the three filters 
used ($H$, S(1), and $K$) were obtained within minutes
of each other, and are therefore suitable for a direct comparison of the colors.

The $H-K$ color of IRS~7 is plotted against the phase of the periodic variability in Fig.~6,
and the average colors in the two major phase ranges are indicated. 
We have divided the color data into two samples: Color during the quiescent phase
from 0.8 to 0.3 (including the ''slow decline'' phase), and during the bright phase from 0.3 to 0.8 phase.
The two-sample Kolmogorov-Smirnov test 
gives only
a 2.5\% probability of these two samples being drawn from the same population.
With a marginal level of significance, the $H-K$ color is redder during the
bright phase.

\begin{figure}
\begin{center}
\figurenum{6}
\includegraphics[scale=0.35,angle=-90]{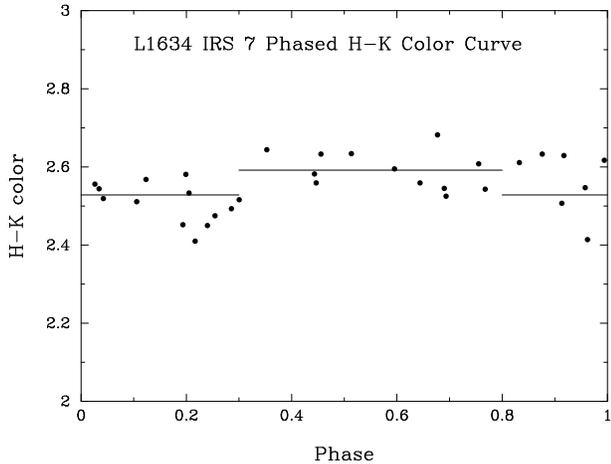}
\caption{
$H - K$ Color curve of L1634 IRS~7 phased with a period of 37.14 days determined from all IRIS and UKIRT data points
by the Lomb-Scargle algorithm. The average color in three phase intervals are indicated. 
There is a small, marginally significant change in color between the quiescent and the outburst phases, with the outburst phase being slightly bluer.
}
\end{center}
\end{figure}

The UKIRT data were precisely aligned using the IRAF task XREGISTER and a small
sub-image containing IRS~7 was selected.
We then identified the light curve phase
of all UKIRT images and after further selecting only UKIRT images with good seeing and
free of other defects, those selected images were arranged in the order of their
phase. Images with similar phases were added together for better signal-to-noise and
form the basis for Fig.~7. In this figure, we show the $K$-band images directly, 
but also have divided the individual phase images
by the average of all low-light images, to normalize the scattering properties of the dust
distribution around the object, and to bring out the spherical nature of the light echoes.

The scattering effects of a light echo can occur in light paths of any inclination
angle $\theta$ against the line of sight, and for such light paths the measured
projected light echo expansion velocity will be $c \cdot sin(\theta)$. In our data, the
brightest individual light echo is seen along what appears to be the wall of the
northern, blueshifted, outflow cavity, and this particular light echo is clealy
moving slower than the projected speed of light, and since it corresponds to the
blueshifted outflow lobe, is is clear that the scattering is at an angle of less
than 90$\arcdeg$ to the line of sight. For a distance measurement, we have to consider
the maximum projected velocity in the overall light echo, and work under the assumption
that this maximum velocity corresponds to 
scattering under a 90$\arcdeg$ angle can. The molecular core L1634 clearly contains
dust, and the young stellar object IRS~7 has formed and is still located at a density
enhancement of this dusty material. There may be extinction effects along different
scattering paths. Our method of normalizing the flux distribution at any given phase
of the variation by the average of the low-light phase is designed to normalize out
effects of uneven illumination or extinction in different scattering paths.
The result of this normalization are light echo images of nearly round appearance,
within the limits of the poor signal-to-noise ratio of those normalized images.
The approximately round outer boundary of these images is therefore assumed to
be light scattered at 90$\arcdeg$ angle, i.e., in the plane of the sky,
since nothing in our data suggests a dust distribution contrary to this assumption.
The fact that the distance so determined is in good agreement
with other, independent, measurements gives added confidence that our
assumption is correct.

\begin{figure}
\begin{center}
\figurenum{7}
\includegraphics[scale=0.5,angle=-90]{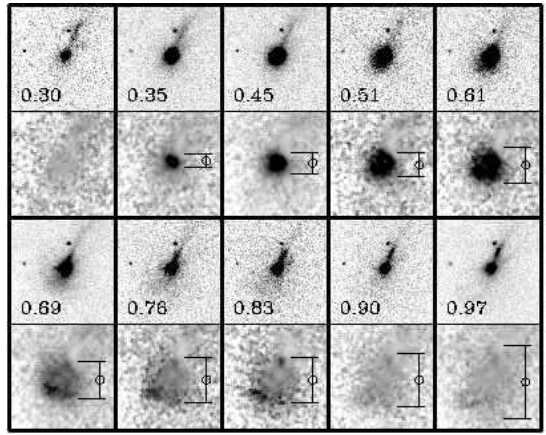}
\caption{
A sequence of phased images and the corresponding normalized light echo images of L1634 IRS~7. 
Each image is $40^{\prime\prime}\times40^{\prime\prime}$ in size.
The available UKIRT $K$ images of L1634~IRS~7 were combined in 
groups of approximately equal phase. The average phase of each image is indicated. 
In order to normalize the scattered light distribution, these
combined images in the bright phase were normalized by the average of all images in the low phase.
The resulting normalized light echo images in the lower half of each phase panel 
show the light echo as an approximately round expanding shell.
}
\end{center}
\end{figure}

We have measured the expansion of the light echoes in the normalized
light echo images by visually fitting a circle around the perimeter
of the normalized light echo. The diameters measured in this rather crude
way are shown as caliper-symbols next to the normalized images in Fig.~7,
to demonstrate that their uncertainties do indeed allow a measurement of the
light echo expansion.
The measurements are plotted in Fig.~8 against the phase of the light curve
together with a linear regression fit and the associate confidence interval of the
resulting distance.
If E is the expansion of 15$\farcs$915 per full period of P = 37.14 d, the
distance in km is:
$d=c \cdot P \cdot 87400 \cdot 206265 / E \equiv 404~pc$.
From the confidence interval of the regression fit, the errors are
$\pm$ 35~pc.
A straight
line with a slope corresponding to the speed of light projected onto the sky
at this distance of 404 pc is shown in Fig. 8.

\begin{figure}
\begin{center}
\figurenum{8}
\includegraphics[scale=0.35,angle=-90]{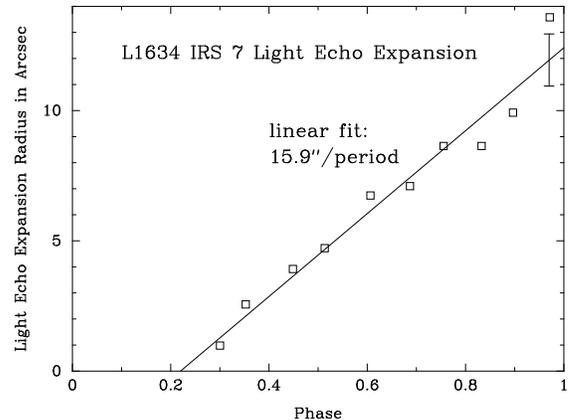}
\caption{
Measured radii of the expanding light echoes of IRS~7, plotted against the phase.
The line regression fit has the slope of the speed of light projected at a distance of 404~pc.
The confidence interval of this fit of $\pm$ 35 pc is indicated as an error bar on the best
fit line.
The non-zero radius measured at the onset of the outburst ($\approx$ phase 0.3) is the result of the
point spread function.
}
\end{center}
\end{figure}

This distance measurement is consistent with the distances to the Orion Nebula
by \citet{Jeffries2007} and \citet{Sandstrom2007}, the arguments presented
by \citet{Bally2009} why L1634 must be physically close to the Orion Nebula,
and with the distance of
400 pc assumed in this paper and by \citet{Bally2009}.
Our measurements are of similar precision to those other distance determinations,
and are entirely independent in their methodology, giving added confidence in
this new consensus distance to the Orion star forming region.

\subsection{Spectroscopy of IRS~7}
The spectra obtained in queue mode with GNIRS on the Fred Gillette Gemini North
telescope were processed using standard procedures in IRAF. 
The raw data were dark subtracted, filtered to reduce pattern noise, and flat-field corrected.
The wavelength calibration is based on Argon lamp spectra. The $K$-band spectra
were extracted using the IRAF routine APPALL, and the resulting spectra were
corrected for telluric absorption by division of an A0 star spectrum
at very nearly identical air mass. 
The strong Br$\gamma$ absorption line in the spectra
of these calibration stars was replaced by a linearly interpolated continuum, so
that we avoid creating artifacts at the nominal position of the Br$\gamma$ line.
Finally, the spectrum was multiplied by
an artificial spectrum of a 9700 K blackbody \citep{Pecaut2013}, to arrive at relative flux units.
Similar procedures were used for the IRTF/SpeX data, 
except that those spectra show more S(1) emission due to the wider slit.

The phase of the spectra was determined from their effective epochs.
It turned out that we got only one GNIRS spectrum near maximum light, three GNIRS
spectra near a phase of 0.90 in the quiescent phase, and the SpeX spectrum at a phase of 0.10,
also in the quiescent phase, as shown by the arrow symbols in Fig.~5.
The three quiescent-phase GNIRS spectra were co-added to improve the signal-to-noise ratio.
All spectral features appear to scale with
the overall broad-band brightness, with the exception of the S(1) line that appears independent
of the variations in IRS~7.

\begin{figure}
\begin{center}
\figurenum{9}
\includegraphics[scale=0.35,angle=-90]{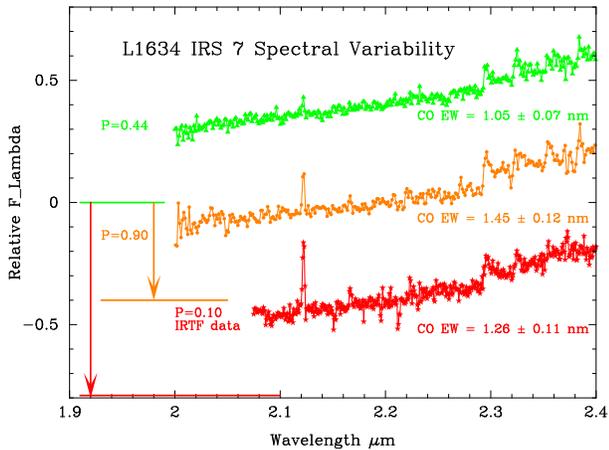}
\caption{
Spectra in the $K$ band, obtained with Gemini/GNIRS and IRTF/SpeX in different phases of the L1634 IRS~7 outbursts.
Only one of the GNIRS spectra (green) was obtained near the maximum of the outburst phase of the light curve.
The SpeX spectrum (red) was obtained during the minimum phase, but shows a higher S/N ratio due to the larger
slit used and longer integration time. The S(1) emission is more prominent in the wider-slit SpeX spectrum, showing
that this emission arises in an extended shock front.
}
\end{center}
\end{figure}

For Fig.~9, the individual spectra were scaled to match their continuum fluxes, and then linearly shifted
to separate them out for the display in the figure. The shift vector is shown in Fig.~8 as an arrow.
We computed the equivalent width of the $^{12}CO$ (2-0) bandhead, the first and best isolated of the CO bandheads.
The continuum was defined as in the CO-index definition by \citet{Kleinmann1986} as the range from 2.2873 - 2.2925 mum, and the line flux
was integrated over the ''line'' wavelength range from 2.2931 - 2.2983 $\mu$m. The errors were computed from the standard deviation
of the flux measurements in the ''continuum'' wavelength range. The resulting equivalent width values
are indicated in Fig. 9. The equivalent widths measured in the quiescent phases are higher than that measured
during the outburst phase, but this effect is not significant at the 3$\sigma$ level. What can be stated
is that the CO-bandhead emission is not constant during the outburst, but it appears not to increase
as strongly as the continuum flux.
Emission of the CO bandheads indicates an optically thin warm, dense inner disk around IRS~7.
The S(1) line emission does not scale with the brightness but is constant and independent
of the continuum flux. We conclude that the S(1) emission
is generated sufficiently far away from IRS~7 that the variations do not affect the
emission of this line.

\subsection{Shock Front Astrometry}
We have studied the proper motions of individual shock-excited emission knots in
the outflows of L1634 IRAS 05173-0555 (the main outflow) and of IRS~7.
The first epoch data were data obtained in 1994 Dec. 8 and 9 with a S(1) line filter in the 
QUIRC infrared camera \citet{Hodapp1996QUIRC} on the UH 2.2m telescope. These data had originally been
published in \citet{Hodapp1995}. For the purpose of this proper motion study,
the data were re-reduced using modern methods. We created separate co-added images
of the western, center, and eastern parts of the L1634 outflow, each aligned using
one reference star common to all the frames in that part of the outflow. The effective
epoch 
of the resulting images was MJD 49695.0 
The second epoch data are the combined images in the H$_2$ 1--0 S(1) filter taken between 
2013 Nov. 7 (MJD 56603.5) and 2014 April 16 (MJD 56760.2) with UKIRT/WFCAM. A total of
27 images were taken over this time span, and the effective epoch of the co-added image
is MJD 56570.2.

The co-added first epoch sub-field images were aligned relative to the combined second epoch
UKIRT images with stars recorded in both images and by using the IRAF GEOMAP and GEOTRAN tasks. This produced
a version of the smaller first epoch image astrometrically matched to the system of astrometric
reference stars in the larger UKIRT second epoch image. In the process, outliers, i.e., stars
with high proper motions, were identified and eliminated. After this prelimary alignment
step, retangular areas were defined covering individual S(1) shock fronts, and the IRAF
task XREGISTER was used to compute the spatial cross-correlation of these rectangular image
sections at both epochs, giving the x and y shift required for maximum correlation. This shift is the proper
motion in pixels. The rectangular areas used are indicated in Fig.~10. The same process was
used on stars in the image. The proper motions measured on stars are generally much
smaller than those in the shock fronts, and give information about the combined
uncertainties caused by reference star proper motion and the typical errors of the cross-correlation
maxima.

\begin{figure}
\begin{center}
\figurenum{10}
\includegraphics[scale=0.25,angle=-90]{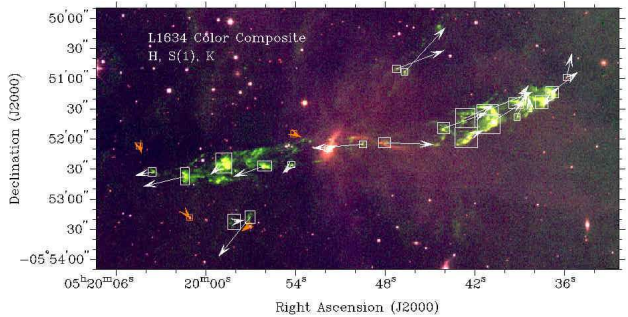}
\caption{
Proper motion of H$_2$ 1--0 S(1) shock fronts measured between 1994 and 2014. The proper motion vectors are scaled to 
represent the extrapolated motion over 1000 years. 
At the distance of 400 pc, 1$\arcsec$ per 1000 yrs corresponds to a projected velocity of 1.9 kms$^{-1}$.
}
\end{center}
\end{figure}

\begin{deluxetable}{ccccc}
\tabletypesize{\scriptsize}
\tablecaption{Shock Front Proper Motions in L1634}
\tablewidth{0pt}
\tablehead{
\multicolumn{2}{c}{RA (J2000) Dec} & \colhead{RA PM} & \colhead{Dec PM} & \colhead{Name}\\
\colhead{{h}\phn{m}\phn{s}} & \colhead{\phn{\arcdeg}~\phn{\arcmin}~\phn{\arcsec}} & \colhead{mas y$^{-1}$} & \colhead{mas y$^{-1}$}
}
\startdata
IRAS & 05173-0555 & Outflow &  \\
05 19 35.9 & $-$5 50 59 & 5 & 23 & HH240E (new)\\
05 19 36.8 & $-$5 51 15 & 25 & 20 & HH240D\\
05 19 37.3 & $-$5 51 22 & 17 & 15 & HH240C\\
05 19 38.3 & $-$5 51 30 & -2 & 12 & HH240B1\\
05 19 39.1 & $-$5 51 24 & 12 & 17 & HH240B2\\
05 19 39.2 & $-$5 51 37 & 10 & 24 & HH240B (SE) \\
05 19 40.7 & $-$5 51 43 & 35 & 22 & HH240A (SW)\\
05 19 40.8 & $-$5 51 32 & 45 & 18 & HH240A (NW)\\
05 19 42.5 & $-$5 51 58 & 17 & 7 & HH240A (SE)\\
05 19 42.7 & $-$5 51 38 & 28 & 13 & HH240A (NE)\\
05 19 44.1 & $-$5 51 50 & 43 & 15 & HH240Z (new)\\
05 19 47.9 & $-$5 52 03 & 48 & -2 & HH240Y (new)\\
05 19 49.6 & $-$5 52 04 & -49 & -4 & HH241V (new)\\
05 19 54.3 & $-$5 52 25 & -8 & -8 & HH241Z (new)\\
05 19 56.1 & $-$5 52 27 & -31 & -10 & HH241A\\
05 19 58.6 & $-$5 52 26 & -13 & -12 & HH241B\\
05 20 01.4 & $-$5 52 37 & -41 & -10 & HH241D\\
05 20 03.5 & -05 52 33 & -17 & -3 & HH241E (new)\\
           &            &   &    \\
IRS~7 & & Outflow & \\
05 19 46.7 & $-$5 50 53 & 40 & 43 & IRS4 (W)\\
05 19 47.2 & $-$5 50 50 & 47 & 18 & IRS4 (E)\\
05 19 57.1 & $-$5 53 18 & -31 & -39 & IRS9 = HH979SE1\\

\enddata
\end{deluxetable}

The two epochs are separated by 6975 days or 19.097 years. In Fig.~10 the measured proper
motion vectors were scaled by 1000/19.097 to represent the extrapolated proper motion
in 1000 years. Table 1 lists the measured proper motion values in units of mas~y$^{-1}$
and labels the individual shock fronts using the nomenclature of \citet{Hodapp1995} for 
the IRS~7 outflow and of \citet{Davis1997} and \citet{OConnell2004} for the IRAS 05173-0555 outflow.
In particular for the latter, the labels in the existing literature were not sufficient to
uniquely identify the shock fronts studied here, so we have assigned additional labels
to some shock fronts, and added descriptive sub-labels to others. In some cases, we had
to resort to adding letter from the end of the alphabet to denote shock fronts upstream
from ''$A$''.
In Fig.~10, we show the measured proper motions of four stars near the L1634 outflows as
-- very short -- arrows in orange color.
The rms uncertainty of the measured proper motions of these stars is 4$\farcs$8 per 1000 years
or 9 km s$^{-1}$ in each coordinate. The typical proper motions of shock fronts are a factor of 10 larger than
this. The reference star system is therefore not the limiting factor in the precision of
our measurements. Proper motions of  bright, well defined shock fronts are highly significant,
while measurements on faint or poorly defined shock fronts are only marginally significant.

\section{Discussion}
\subsection{The Variability of IRS~7}

Young stars are generally variable, in particular if high-precision observations
with space-based telescopes are considered.
For young stars sufficiently evolved to be optically detectable,
the variable star T Tauri is the prototypical example for the small,
irregular variability observed in virtually all of them.
Based on optical photometry of a large sample of T Tauri stars, \citet{Herbst1994} 
identified three types of variability: rotationally modulated cool star spots,
variable accretion leading to changes in the veiling observed in some stars,
and variable circumstellar obscuration.

Variability surveys of star-forming regions at infrared wavelengths, 
primarily with the warm $Spitzer$ space
telescope, have expanded our knowledge of young star variability to
objects sufficiently young and embedded to only be detectable at
infrared wavelengths.
The Orion nebula region was monitored in the $Spitzer$ IRAC channels 1 and 2 by
the Young Stellar Object Variability (YSOVAR) project, and \citet{Morales2011} have published examples
of the lightcurves obtained in their month-long high-cadence monitoring
campaign. The one star among their published light curves that has some
resemblance to IRS~7 is the Class I object 2M J053435.44-053959.1, which
they describe as being caused by slow changes of accretion over the 40 day
time span of the observations. 

Variability dominated by accretion bursts was studied in NGC2264 by \citet{Stauffer2014} and \citet{Cody2014}
using YSOVAR $Spitzer$ data.
However, these accretion burst stars are characterized by shorter, more frequent, and
non-periodic outbursts.
The few examples in their
sample with significant periodicity have periods in the range of 7 days.
It should be noted that their whole observing campaign of NGC2264 lasted
only about 40 days, and a procedure for subtracting long-term trends in the
light curves was used, so that this study has a reduced sensitivity
to variations on timescales longer than $\approx$ 10 days, and would not
be able to recognize periodicity with periods of 37 days, such as IRS~7 in
L1634. In summary, the results by 
\citet{Stauffer2014}
and \citet{Cody2014} show that outburst phenomena are frequently observed
in YSOs on timescales shorter than those observed in IRS~7.

On the other side, more substantial outbursts of long duration and
larger amplitude in young, still accreting stars
are traditionally classified into either FU Orionis (FUor) or EX Lupi (EXor) type
outbursts, depending on the duration of the outburst and its spectrum.
The first review of these phenomena has been given by \citet{Herbig1977}.
FUor outbursts exhibit time scales of decades to centuries and the spectral
characteristics of an optically thick luminous disk with absorption line spectra,
including CO bandhead absorption \citep{Aspin2009}.
The less substantive EXor outbursts have timescales of years, have
been observed by \citet{Aspin2006,Aspin2010} to actually return to pre-outburst brightness, and show
optically thin emission line spectra, in particular, CO bandhead emission.
In some cases, e.g.,
V1647 Ori, repetitive outburst have been observed \citep{Aspin2006}.

FUor and EXor outbursts are rare phenomena and most known cases have
been discovered by optical methods where large surveys and monitoring
campaigns had been feasible in the recent past. A small number of more
deeply embedded outburst objects have now been found that are only detectable
at infrared wavelengths. Most of these cannot be clearly classified as
FUor or EXor outbursts, but exhibit outburst amplitudes, durations,
and spectral characteristics different from either of those two established
classes. One example is OO~Ser \citep{Hodapp1996,Hodapp1999,Hodapp2012} that
showed an outburst amplitude of only 2.5 mag and a duration of the fast part
of the outburst of only a few years, but has not settled back to its pre-outburst
brightness in the two decades since its outburst. 
An outburst of an even more deeply embedded object, HOPS 383, a Class 0 protostar, was
recently reported by \citet{Safron2015}, showing that accretion instabilies can occur
very early in the evolution of a protostar, when the object emits most of its
energy at far-infrared and sub-mm wavelengths.

\citet{Hillenbrand2015} have recently discussed the various outburst
phenomena in the context of determining the rate of FUor outburst,
the highest amplitude and longest duration extreme of the distribution
of outburst events. From this, and many papers discussing individual
objects, it is clear that outburst events in young stars span timescales
from hours to millennia, and burst amplitudes ranging from a few percent
to several orders of magnitude in flux. The amplitude and period of IRS~7
fits right into this distribution, but the well-established periodicity
makes it stand out among otherwise similar objects. It is not unique
in this respect, however. For example, \citet{Hodapp2012} reported on
the periodic (543d) variability of the illuminating star V371 Ser of the "cometary"
nebula EC 53 in Serpens NW, a very young object associated with an
H$_2$ jet and showing other signs of extreme youth.

For IRS~7, the fact that light echoes can be observed to propagate into
the surrounding dusty material (Fig.~7) with the speed of light at the
distance of 404 pc argues against any effect that is specific
to the line of sight towards the observer, e.g., an eclipsing binary or
periodic obscurations by a tilted disk or orbiting dust cloud. Also, the
near color neutrality of the variation (Fig.~6) speaks against a dust obscuration effect.

Rotational modulation of an accretion hot spot at first seems a plausible
explanation of the IRS~7 variability, since it spends about have of the
time in the bright state, and the other half at nearly constant, low light.
However, the light echoes appear round and are centered on the illuminating
object. A hot spot on a rotating star or disk would be expected to form
alternating light echoes on either side of the star, if the scattering is
close to isotropic.

The only spectral feature clearly detected in our $K$-band spectra are the
CO bandheads in emission and emission from shock-excited H$_2$. 
The shock excited S(1) line flux appears independent of the phase of the
variability, while the CO bandhead emission scales approximately with the
overall continuum flux variations.

The very young evolutionary state of IRS~7
as the driving source of a bipolar outflow suggests that accretion luminosity
in a warm disk is the main source of luminosity. The fact that the 2.29 $\mu$m
CO bandheads are seen in emission indicates that the warm inner parts of the accretion
disk are optically thin, and in this respect IRS~7 resembles an EX Lupi eruptive variable \citep{Aspin2010}.
The CO bandhead emission at 2.29 $\mu$m is present both in the quiescent and the outburst phase,
indicating that an optically thin warm disk is always present. The equivalent width of the
$^{12}CO (2-0)$ emission is slightly smaller in the outburst phase than in the quiescent phase,
indicating the the CO emitting part of the disk is not primarily responsible for the
increase in luminosity during the outburst. We caution that this result is statistically
only marginally significant. The CO line emission is stronger during outburst, but
not quite by the same factor as the continuum emission.

It is therefore reasonable to assume that the variability
is caused by instabilities in the accretion process and that the warm gas responsible
for the CO bandhead emission never completely disappears, even during the low accretion
phase of the light curve. 

A short period
binary companion is a possible explanation for the periodic accretion instabilities.
This mechanism of binary orbit triggered accretion instability was discussed
by \citet{Artymovicz1996} and modeled in detail by \citet{Val-Borro2011}.
While possible, this binary-triggered accretion model has the disadvantage
of postulating the existence of a close binary companion, which we have no
other evidence for.
Assuming a mass of 0.5 \msun for IRS~7, a 37 d = 0.10 y orbital period has
a 0.11 AU semi-major axis. Such a close binary orbiting within the disk feeding
the accretion would, most likely, disrupt that disk in a short time.
Thus we conclude that a short period binary companion is an unlikely explanation for the periodic accretion.

We prefer a model where the accretion instabilities are intrinsic
to the accretion process itself, and do not require an external trigger.
A theoretical framework has recently been developed by \citet{DAngelo2010}
for magnetosphere-disk interactions in the case of a star with strong dipolar 
magnetic field surrounded by a thin accretion disk in Keplerian rotation.
They find that under certain conditions, accretion stalls out and matter
accumulates in the disk just outside of the co-rotation radius. When the 
gas pressure in that accumulated material becomes high enough, it overcomes
the centrifugal barrier created by the interaction between the magnetic field
and the disk, and accretion resumes. After the material has been transferred
to the star, the inner disk edge will move back outside the co-rotation radius,
and the stage is set for a repetition of this process. This mechanism provides a
plausible explanation for cyclical accretion onto a young star.
In a follow-on paper, \citet{DAngelo2012} specifically apply their model to
the repetive eruptions of the prototypical EXor EX Lupi. They make the point
that in addition to the major outbursts, ''higher frequency oscillations with a period
of about 30 d and amplitude of 1 -- 2 mag'' are observed. These higher frequency 
oscillations are very similar to the variations observed in L1634 IRS~7.

\subsection{Shock Front Proper Motions}

The comparison of the 1994 UH/QUIRC and 2014 UKIRT/WFCAM data clearly shows proper motion of the
faint emission knots to the NW and SE of IRS~7 -- infrared sources 4 and 9 in the nomenclature of \citet{Hodapp1995} --
consistent with them originating
from IRS~7. The measured tangential velocities are 
112.3 km s$^{-1}$ for IRS~4 (NW of IRS~7) and
94.1 km s$^{-1}$ for IRS~9 4 (SE of IRS~7).

From the fact that we see near-infrared radiation, presumably scattered light, to
the NW of IRS~7, but not in the opposite direction, we conclude that the NW lobe of the bipolar
outflow from IRS~7 is oriented towards the observer, i.e., blueshifted.
We detect the HH 979 NW1 shock front found by \citet{Bally2009}, but do not have first epoch
S(1) images to determine a proper motion. We do not detect any S(1) emission from HH 979 NW2, where
\citet{Bally2009} found another [SII] shock front. This could be explained by this farthest shock
front having broken out of the L1634 molecular cloud through its front surface.

We also measured the proper motions of many of the shock fronts and emission knots
in the main outflow originating in IRAS 05173-0555. In particular, we measure proper motions in opposite direction
for the two small shock fronts closest to the position of the IRAS source, proving that the
main outflow does indeed originate from this deeply embedded protostar. We detect 
scattered continuum light in the $K$ band (red in Figs. 1 and 10) from the inner walls of the outflow
cavity of IRAS 05173-0555. The driving source itself, or the innermost parts of the outflow cavity
in the direction of the blueshifted western outflow lobe, is seen as an unresolved continuum
emission source in our WFCAM $K$-band image, while it is undetected in the $H$ band and only
marginally detected in the narrow-band S(1) image. This continuum source is also located
at the apex of the thin lines of continuum emission that apparently outline the walls of the
outflow cavity in the west. A similar outline of the eastern outflow cavity walls is also visible,
but cannot be traced all the way to the unresolved source, due to absorption from the
nearly edge-on disk surrounding the driving source. The point source, presumably at or very
near the driving protostellar source of the outflow, is located at 
$\alpha(2000)$ = 05:19:48.37, $\delta(2000)$ = -05:52:03.9,
consistent with the positions given by \citet{Beltran2002} for the sub-mm source, and by
\citet{Beltran2008} for the peak of the NH$_3$ position in VLA measurements.

The two innermost S(1) shock fronts in the IRAS 05173-0555 outflow are not equidistant from this
position of the driving source. The eastern shock front is 18$\arcsec$ east of the
central source, while the western shock front is only 6\farcs3 from that source.
The mid-point between these innermost shock fronts -- located at
$\alpha(2000)$ = 05:19:48.74 -- does not 
show any detectable $K$-band continuum source, is not morphologically consistent with the
apex of the cavity outline, and does not coincide with any significant feature in the maps
of \citet{Beltran2002} and \citet{Beltran2008}.

The projected velocity of the eastern shock is 94.3 km~s$^{-1}$ while the western shock
is 93.0 km~s$^{-1}$. Such high velocities would dissociate the H$_2$ molecule if the shock
were against stationary gas. We conclude that both shocks are internal shocks with much
lower relative shock velocity in a wind moving with $\approx$ 90 km~s$^{-1}$. Exactly where
detectable S(1) emission is produced depends on the details of the internal velocity structure
of the wind, and is plausibly different in the two lobes.

\citet{OConnell2004} measured radial velocities in both the blue-shifted (HH240) and red-shifted (HH241)
lobe of the IRAS 05173-0555 outflow, relative to the systemic velocity of +8 km s$^{-1}$ obtained
by \citet{DeVries2002}. The radial velocities cover a broad range, but are generally centered
around 30 km s$^{-1}$.
Compared to the tangential velocities derived here from the astrometry of order of 90 km s$^{-1}$,
this suggests that this outflow
is oriented only about 18$\arcdeg$ from the plane of the sky, consistent with the morphology.

While \citet{Bally2009} found the HH 980 system of Herbig-Haro objects associated with
L1634N in their [SII] image, we do not detect any H$_2$ 1--0 S(1) emission from this
YSO. There is some S(1) emission detectable in that general area, but this appears to
be the fluorescently excited outer layer of the molecular cloud and associated with 
the photo-dissociation of H$_2$ by the interstellar radiation field.
Most of the L1634N outflow propagates outside of the bright rim of L1634, in a region 
devoid of H$_2$.

\section{CONCLUSIONS}
The embedded outflow source IRS~7 in the L1634 molecular cloud was
found to experience periodic outbursts in brightness with a period
of $37.14 \pm 0.04$ days and a typical peak-to-peak amplitude of 2 mag.
Both the quiescent light brightness and the outburst amplitude are
changing on the level of 0.5 mag over the monitoring period in a
apparently random fashion with time constants many times larger than
the main 37.14 day period. 
The outbursts are associated with only small changes in the $H-K$ color,
and we do not observe significant changes in the shape of the spectrum or
individual spectral features.
The rapid outbursts produce light echoes propagating into
the surrounding molecular cloud at the projected speed of light, 
demonstrating that the brightness variations of IRS~7 are
due to a mechanism that illuminates the surrounding dusty cloud isotropically.
The expansion of the light echoes allows the distance to L1634 to be
measured as d = 404 pc $\pm$ 35 pc.
We have used images in the S(1) line in comparison with similar data
obtained in 1994 to measure the proper motions of the S(1) shock fronts
in the two southern outflows. This confirms that there are indeed two
independent outflows, the larger one powered by the FIR source
IRAS 05173-0555, the other by IRS~7.

\acknowledgments

Most photometric data on L1634 IRS~7 were obtained at the IRIS telescope of the
Universit\"atssternwarte Bochum on Cerro Armazones,
which is operated under a cooperative agreement between 
the "Astronomisches Institut, Ruhr Universit\"at Bochum", Germany, the "Universidad Catolica del Norte" in Antofagasta, Chile,
and the Institute for Astronomy, University of Hawaii, USA. 
The operation of the IRIS telescope is supported by the ``Nordrhein-Westf\"alische Akademie der
Wissenschaften und der K\"unste'' in the framework of the academy program by the Federal
Republic of Germany and the state of Nordrhein-Westfalen.
Construction of the IRIS infrared
camera was supported by the National Science Foundation under grant AST07-04954.
We wish to thank 
Angie Barr Dominguez,
Thomas Dembsky, 
Holger Drass, 
Francisco P. Nunez,
Lena Kaderhandt, 
Michael Ramolla 
and 
Christian Westhues
for
operating the IRIS telescope for the acquisition of the data used in this paper,
Ramon Watermann for writing the data reduction pipeline,
and Roland Lemke for technical support.

When the UKIRT data for this project were acquired, UKIRT was operated by the Joint
Astronomy Centre on behalf of the Science and Technology Facilitees Council of the United Kingdom.

The spectroscopy was mostly obtained at the Gemini Observatory, which is operated by the 
Association of Universities for Research in Astronomy, Inc., under a cooperative agreement 
with the NSF on behalf of the Gemini partnership: the National Science Foundation 
(United States), the National Research Council (Canada), CONICYT (Chile), the Australian 
Research Council (Australia), Minist\'{e}rio da Ci\^{e}ncia, Tecnologia e Inova\c{c}\~{a}o 
(Brazil) and Ministerio de Ciencia, Tecnolog\'{i}a e Innovaci\'{o}n Productiva (Argentina).

One spectrum was obtained at the Infrared Telescope Facility (IRTF), which 
is operated by the University of Hawaii under contract NNH14CK55B
with the National Aeronautics and Space Administration. We thank J. S. Bus, M. Connelley,
and J. T. Rayner for their support during the observations.

This publication makes use of data products from the Two Micron All Sky Survey, which is
a joint project of the University of Massachusetts and the Infrared Processing and Analysis Center/
California Institute of Technology, funded by the National Aeronautics and Space Administration
and the National Science Foundation.

{\it Facilities:} \facility{Gemini:Gillett(GNIRS)},
\facility{IRTF(SpeX)},
\facility{IRIS},
\facility{UKIRT(WFCAM)}

\clearpage



\begin{thebibliography}{}


\bibitem[Artymovicz \& Lubow(1996)]{Artymovicz1996}
Artymovicz, P. \& Lubow, S. H. 1996, \apj, 467, L77

\bibitem[Aspin et al.(2006)]{Aspin2006}
Aspin, C., Barbieri, C., Boschi, F., et al. 2006, \aj, 132, 1298

\bibitem[Aspin et al.(2009)]{Aspin2009}
Aspin, C., Greene, T. P., \& Reipurth 2009, \aj, 137, 2968

\bibitem[Aspin et al.(2010)]{Aspin2010}
Aspin, C., Reipurth, B., Herczeg, G. J., \& Capak, P. 2010, \apj, 719, 50

\bibitem[Bally et al.(2009)]{Bally2009}
Bally, J., Walawender, J., Reipurth, B., \& Megeath, S. T. 2009, \aj, 137, 3843

\bibitem[Beltr\'{a}n et al.(2002)]{Beltran2002}
Beltr\'{a}n, M. T.,
Estalella, R.,
Ho, P. T. P.,
Calvet, N.,
Anglada,G.,
\& Sep\'{u}lveda, I.
2002, \apj, 565, 1069

\bibitem[Beltr\'{a}n et al.(2008)]{Beltran2008}
Beltr\'{a}n, M. T.,
Wiseman, J.,
Ho, P. T. P.,
Estalella, R.,
Fuller, G. A.,
\& Wootten, A.
2008, \aap, 485, 517


\bibitem[Bertin(2005)]{Bertin2005}
Bertin, E. 2005, in Astronomical Data Analysis Software and Systems XV,
ASP Conference Series, Vol. 351, eds. Gabriel, C., Arviset, C, Ponz, D., \& Solano, E., 112

\bibitem[Bertin \& Arnouts(1996)]{Bertin1996}
Bertin, E. \& Arnouts, S. 1996, \aaps, 117, 393

\bibitem[Casali et al.(2007)]{Casali2007}
Casali, M., Adamson, A., Alves de Oliveira, C. et al. 2007, \aap, 467, 777

\bibitem[Chini et al.(1997)]{Chini1997}
Chini, R., Reipurth, B., Sievers, A., Ward-Thompson, D, Haslam, C. G. T.,
Kreysa, E., \& Lemke, R. 1997, \aap, 325, 542

\bibitem[Cody et al.(2014)]{Cody2014} Cody, A.~M., Stauffer, J., 
Baglin, A., et al.\ 2014, \aj, 147, 82 

\bibitem[Cohen, Harvey, \& Schwartz(1985)]{Cohen1985}
Cohen, M., Harvey, P. M., \& Schwartz, R. D. 1985, \apj, 296, 633

\bibitem[D'Angelo \& Spruit(2010)]{DAngelo2010}
D'Angelo, C. R. \& Spruit, H. C. 2010, \mnras, 406, 1208

\bibitem[D'Angelo \& Spruit(2012)]{DAngelo2012}
D'Angelo, C. R. \& Spruit, H. C. 2012, \mnras, 420, 416

\bibitem[Davis et al.(1997)]{Davis1997}
Davis, C. J., Ray, T. P., Eisl\"offel, J., \& Corcoran, D. 1997, \aap, 324, 263

\bibitem[de Val-Borro et al.(2011)]{Val-Borro2011}
de Val-Borro, M., Gahm, G. F., Stempels, H. C., \& Peplinski, A. 2011,
\mnras, 413, 2679

\bibitem[de Vries, Narayanan, \& Snell(2002)]{DeVries2002}
de Vries, C. H., Narayanan, G., \& Snell, R. L. 2002, \apj, 577, 798

\bibitem[Elias et al.(2006)]{Elias2006}
Elias, J. H., Joyce, R. R., Liang, M., Muller, G. P., Hileman, E. A., \& George, J. R. 2006,
in Ground-based and Airborne Instrumentation for Astronomy, eds. I. S. McLean and I. Masanori,
SPIE, 6269, 138


\bibitem[Herbig(1977)]{Herbig1977}
Herbig, G.~H.\ 1977, \apj, 217, 693

\bibitem[Herbst et al.(1994)]{Herbst1994}
Herbst, W., Herbst, D. K., Grossman, E. J., \& Weinstein, D. 1994, \aj, 108, 1906

\bibitem[Hillenbrand \& Findeisen(2015)]{Hillenbrand2015}
Hillenbrand, L. A. \& Findeisen, K. P. 2015, \apj, 808, 68

\bibitem[Hodapp(1994)]{Hodapp1994}
Hodapp, K.-W. 1994, \apjs, 94, 615

\bibitem[Hodapp(1999)]{Hodapp1999}
Hodapp, K. W. 1999, \aj, 118, 1338

\bibitem[Hodapp et al.(2010)]{Hodapp2010}
Hodapp, K. W., Chini, R., Reipurth, B., Murphy, M., Lemke, R.,
Watermann, R., Jacobson, S., Bischoff, K., Chonis, T., Dement, K.,
Terrien, R., \& Provence, S. 2010, Proc. SPIE 7735-45.

\bibitem[Hodapp et al.(2012)]{Hodapp2012} Hodapp, K.~W., Chini, 
R., Watermann, R., \& Lemke, R.\ 2012, \apj, 744, 56 

\bibitem[Hodapp et al.(1996)]{Hodapp1996QUIRC}
Hodapp, K.-W., Hora, J.~L., Hall, D.~N.~B., et al.\ 1996, 1, 177 

\bibitem[Hodapp et al.(1996)]{Hodapp1996} Hodapp, K.-W., Hora, 
J.~L., Rayner, J.~T., Pickles, A.~J., \& Ladd, E.~F.\ 1996, \apj, 468, 861 

\bibitem[Hodapp \& Ladd(1995)]{Hodapp1995}
Hodapp, K.-W. \& Ladd, E. F. 1995, \apj, 453, 715

\bibitem[Hodgkin et al.(2009)]{Hodgkin2009}
Hodgkin, S. T., Irwin, M. J., Hewett, P. C., \& Warren, S. J. 2009, \mnras, 394, 675

\bibitem[Jeffries(2007)]{Jeffries2007} 
Jeffries, R. D. 2007, \mnras, 376, 1109

\bibitem[Jones et al.(1984)]{Jones1984} 
Jones, B. F., Cohen, M., Sirk, M., Jarrett, R. 1984, \aj, 89, 1404

\bibitem[Kleinmann \& Hall (1986)]{Kleinmann1986}
Kleinmann, S. G. \& Hall, D. N. B. 1986, \apjs, 62, 501

\bibitem[Lee et al.(2000)]{Lee2000}
Lee, C.-F., Mundy, L. G., Reipurth, B., Ostriker, E. C., \& Stone, J. M. 2000,
\apj, 542, 925

\bibitem[Lomb(1976)]{Lomb1976}
Lomb, N. R. 1976, Ap\&SS, 39, 447

\bibitem[Morales-Calder{\'o}n et al.(2011)]{Morales2011} 
Morales-Calder{\'o}n, M., Stauffer, J.~R., Hillenbrand, L.~A., et al.\ 
2011, \apj, 733, 50 

\bibitem[O'Connell et al.(2004)]{OConnell2004}
O'Connell, B., Smith, M. D., Davis, C. J., Hodapp, K. W., Khanzadyan, T., \& Ray, T. 2004,
\aap, 419, 975

\bibitem[Pecaut \& Mamajek(2013)]{Pecaut2013}
Pecaut, M. J. \& Makajek, E. E. 2013, \apjs, 208, 1

\bibitem[Rayner et al.(2003)]{Rayner2003}
Rayner, J. T., Toomey, D. W., Onaka, P. M., Denault, A. J., Stahlberger, W. E., Vacca, W. D.,
Cushing, M. C., \& Wang, S. 2003, PASP, 115, 363

\bibitem[Reipurth et al.(1993)]{Reipurth1993}
Reipurth, B., Chini, R., Kr\"ugel, E., \& Sievers, A. 1993, \aap, 273, 221

\bibitem[Safron et al.(2015)]{Safron2015} Safron, E.~J., Fischer, 
W.~J., Megeath, S.~T., et al.\ 2015, \apjl, 800, L5 

\bibitem[Sandstrom et al.(2007)]{Sandstrom2007}
Sandstrom, K. M., Peek, J. E. G., Bower, G. C., Bolatto, A. D., \& Plambeck, R. L. 2007,
\apj, 667, 1161

\bibitem[Scargle(1982)]{Scargle1982}
Scargle, J. D. 1982, \apj, 263, 835

\bibitem[Skrutskie et al.(2006)]{Skrutskie2006}
Skrutskie, M. F., Cutri, R. M., Stiening, R., Weinberg, M. D., Schneider, S.,
Carpenter, J. M., Beichman, C., Capps, R., Chester, T., Elias, J., Huchra, J.,
Liebert, J., Lonsdale, C., Monet, D. G., Price, S., Seitzer, P., Jarrett, T.,
Kirkpatrick, J. D., Gizis, J., Howard, E., Evans, T., Fowler, J., Fullmer, L., 
Hurt, R., Light, R., Kopan, E. L., Marsh, K. A., McCallon, H. L., Tam, R.,
Van Dyk, S., \& Wheelock, S. 2006, \aj, 131, 1163

\bibitem[Stauffer et al.(2014)]{Stauffer2014} Stauffer, J., Cody, 
A.~M., Baglin, A., et al.\ 2014, \aj, 147, 83 


\bibitem[Tody(1986)]{Tody1986}
Tody, D. 1986, in Proc. SPIE, Instrumentation in Astronomy VI,
ed. D. L. Crawford, 627, 733

\bibitem[Tokunaga, Simons, \& Vacca(2002)]{Tokunaga2002}
Tokunaga, A. T., Simons, D. A., \& Vacca, W. D. 2002, \pasp, 114, 792

\end{thebibliography}
\end{document}